\documentclass[preprint,12pt]{elsarticle}

\usepackage{amssymb}

\journal{Nuclear Physics B}

\begin{document}

\begin{frontmatter}



\title{Criticality of
the $(2+1)$-dimensional $S=1$ transverse-field Ising model
with extended interactions:
Suppression of corrections to scaling}


\author{Yoshihiro Nishiyama}

\address{Department of Physics, Faculty of Science,
Okayama University, Okayama 700-8530, Japan}

\begin{abstract}

The criticality of the $(2+1)$-dimensional $S=1$
transverse-field Ising model
is investigated with
the numerical diagonalization method.
The scaling behavior is improved by tuning the
coupling-constant parameters;
the $S=1$ spin model allows us to incorporate
a variety of interactions
such as the single-ion anisotropy and the biquadratic interactions.
Simulating the clusters with $N=6,8,\dots,20$ spins,
we estimate
the critical indices as $1/\nu=1.586(6)$ and $\eta=0.034(4)$.
\end{abstract}


\end{frontmatter}



\section{\label{section1} Introduction}

The finite-size-scaling method allows us to estimate the critical indices
from the simulation data.
In practice, however, there appear
sub-leading singularities, 
the so-called corrections to scaling.
Corrections to scaling are suppressed gradually
by enlarging the system sizes.
On the one hand, there was reported a scheme to control corrections to scaling
directly by tuning the coupling-constant parameters 
\cite{Chen82,Symanzik83a,Symanzik83b,Hasenfratz94,Hasenfratz98,Blote95,Fernandez94,Ballesteros98,Hasenbusch99,Hasenbusch00}.
Particularly, in the lattice-field theory, the lattice artifact brings 
about a severe obstacle to extracting information through the continuum 
limit. There have been reported a number of attempts to suppress the lattice 
artifact by tuning extended interaction parameters; see Ref. \cite{Hasenfratz98}
for a review. This issue is closely related to the analysis of criticality,
because the critical-point phenomena are also governed by long-wave-length 
fluctuations. In practice, it is even desirable to develop a scheme \cite{Fernandez94},
where no {\it ad hoc} adjustment of interaction parameters is required.
This approach may be particularly of use
for the numerical diagonalization method, with which
the tractable system size is 
restricted severely.
(On the contrary, as for the Monte Carlo method,
it might be more rewarding to enlarge the system size
rather than incorporate extra interactions.)

Motivated by the above idea, in Ref. \cite{Nishiyama06},
we investigated the
three-dimensional (classical)
Ising model with the extended (next-nearest-neighbor and plaquette-four-spin) interactions
by means of the diagonalization (transfer-matrix) method for the clusters with
$N \le 13$ spins.
Thereby, we estimated
the critical indices as $\nu=0.6245(28)$ and $y_h=2.4709(73)$.
In the present paper, based on this preliminary survey,
we shall make the following improvements.
First, we consider the quantized version of the Ising model,
namely, the $(2+1)$-dimensional transverse-field Ising model.
(Note that the $(2+1)$-dimensional ground-state phase transition
belongs to the same universality class as that of the
three-dimensional classical counterpart.)
The quantum Hamiltonian matrix is sparse (the matrix has few non-zero elements),
as compared to the transfer matrix of the classical model.
Taking this advantage, we treat the system sizes up to $N \le 20$.
Second, we extend the magnitude of constituent spin to $S=1$ from $S=1/2$.
Owing to this extension, we are able to
incorporate a variety of interactions
such as the single-ion anisotropy $D$ and the biquadratic interactions, $J_4$ and $J_4'$.
To be specific, we consider the Hamiltonian
\begin{equation}
\label{Hamiltonian}
{\cal H}= 
-J \sum_{\langle ij \rangle} S^z_i S^z_j
- J'\sum_{\langle \langle ij \rangle \rangle} S^z_i S^z_j
- J_4 \sum_{\langle ij \rangle} (S^z_i S^z_j)^2
- J_4' \sum_{\langle \langle ij \rangle \rangle} (S^z_i S^z_j)^2
  +D \sum_{i} (S^z_i)^2
     -\Gamma \sum_i S^x_i
              .
\end{equation}
Here, the quantum $S=1$ operators
$\{ {\bf S}_i \}$ are placed at each 
square-lattice point $i$.
The summations, 
$\sum_{\langle ij \rangle}$ and
$\sum_{\langle \langle ij \rangle \rangle}$,
run over all possible nearest-neighbor and next-nearest-neighbor pairs,
respectively.
The parameters $J$ and $J'$ are the corresponding coupling constants.
The parameter $\Gamma$ denotes  
the transverse magnetic field.   
Last, 
these coupling-constant parameters are adjusted to
\begin{eqnarray}
(J,J',J_4,J_4',D) 
&=& (
 0.4119169708 5 ,  
0.1612506961 6 ,  
-0.1176402001 8 ,    \nonumber \\  
\label{fixed_point}
& & 
-0.0 526792660 1 ,  
-0.3978195612 2   
)
.
\end{eqnarray}
This finely adjusted parameter set renders a significant improvement to the scaling
behavior.
We determine the parameter set 
in Sec. \ref{section2}, based on a real-space decimation (Fig. \ref{figure1}).
The $\Gamma$-driven phase transition is our concern (Sec. \ref{section3}).

In fairness, it has to be mentioned that the Ising model
with $S=1$ and $D=\ln2$
was investigated \cite{Deng03}  by means of the Monte Carlo method.
At $D=\ln 2$, the cluster-update algorithm works efficiently.
We stress that in our simulation, the anisotropy $D$ is tuned so as to suppress the
finite-size errors; technically, 
the value of $D$ is not subjected to any restriction
in the 
numerical-diagonalization scheme.

The rest of this paper is organized as follows.
In Sec. \ref{section2},
setting up a real-space-decimation scheme,
we determine a (non-trivial) fixed point (\ref{fixed_point}). 
In Sec. \ref{section3}, around the fixed point,
we perform a large-scale computer simulation; the simulation scheme is explicated in the
Appendix.
Taking the advantage of suppressed scaling errors,
we analyze the criticality of the $d=3$ Ising universality class
with the diagonalization method.
In Sec. \ref{section4},
we preset the summary and discussions, referring to related studies.

\section{\label{section2}
Search for a scale-invariant point: 
Suppression of finite-size corrections}

As mentioned in the Introduction,
the coupling-constant parameters,
$(J,J',J_4,J_4',D)$,
are set to
Eq. (\ref{fixed_point}).
In this section, we explain how 
this parameter set is determined.
%
Here, we stress that the analysis (real-space decimation)
of this section is not intended to determine the critical exponents.
The critical exponents are determined in the
subsequent scaling analysis in Sec. \ref{section3}.

\begin{figure}
\includegraphics[width=100mm]{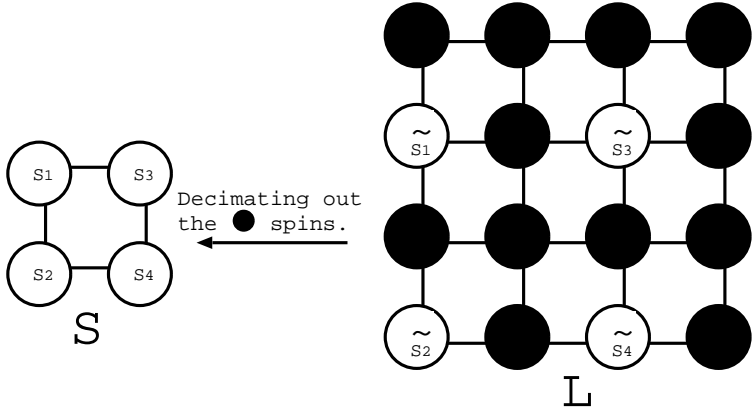}%
\caption{\label{figure1}
A schematic drawing of the real-space renormalization group (decimation)
for the $(2+1)$-dimensional transverse-field Ising model 
(\ref{Hamiltonian}) is presented.
Through decimating out the spin variables indicated by the symbol $\bullet$ within the 
$L$ cluster, we obtain a coarse-grained lattice identical to the
$S$ cluster.
Imposing the scale-invariance conditions, Eqs. 
(\ref{scale_invariance1})-(\ref{scale_invariance5}),
we arrive at the fixed point (\ref{fixed_point}).
}
\end{figure}

To begin with, we outline the underlying idea of 
the finely-tuned parameters, Eq. (\ref{fixed_point}).
The parameter set (\ref{fixed_point}) is a scale-invariant (fixed) point of a real-space decimation.
In Fig. \ref{figure1},
we present a schematic drawing of the
real-space decimation for a
couple of clusters labeled by $S$ and $L$.
The cluster sizes are
$2\times 2$ and $4\times 4$, respectively.
As indicated, through 
a decimation of the $\bullet$ spins,
the $L$ cluster
reduces to
a coarse-grained one identical to the $S$ cluster.
Our aim is to search for a scale-invariant point with respect to the real-space
decimation.

Before going into the fixed-point analysis,
we set up the simulation scheme for the clusters $S$ and $L$.
We cast the Hamiltonian into the following plaquette-based
expression
\begin{equation}
\label{plaquette_based}
{\cal H}= \sum_{ [ijkl] } {\cal H}^\Box_{ijkl}
               +D\sum_i (S^z_i)^2 - \Gamma \sum_i S^x_i 
          ,
\end{equation}
with the plaquette interaction
\begin{eqnarray}
{\cal H}^\Box_{ijkl} & = &
- \frac{J}{2}(S^z_i S^z_j +S^z_jS^z_l+S^z_kS^z_l+S^z_iS^z_k)
   -J' (S^z_iS^z_l+S^z_jS^z_k)                                        \nonumber    \\
   & & - \frac{J_4}{2}((S^z_i S^z_j)^2 +(S^z_jS^z_l)^2   
    + (S^z_kS^z_l)^2+(S^z_iS^z_k)^2)   \nonumber \\
 & &  -J_4' ((S^z_iS^z_l)^2+(S^z_jS^z_k)^2)    .
\end{eqnarray}
(The denominator $2$ compensates
the duplicated sum.)
Hence, the Hamiltonian for the $S$ cluster is given by
\begin{equation}
{\cal H}_S=(1+b){\cal H}_{1234}^{\Box}  +D\sum_{i=1}^4 (S^z_i)^2 
-\Gamma\sum_{i=1}^4 S^x_i    .
\end{equation}
Here, the parameter $b$ controls the boundary-interaction strength,
and hereafter, we set $b=0.75$;   
we consider the validity of this choice afterward.
The boundary-interaction parameter $b$ interpolates smoothly the
open ($b=0$) and periodic ($b=1$)
boundary conditions.
The point is that for such a small cluster with the linear dimension $L=2$,
the bulk and boundary interactions are indistinguishable.
Such a redundancy is absorbed into the tunable parameter $b$;
in other worlds, the parameter $b$ 
is freely tunable without violating the
translation invariance.
We make use of this redundancy to 
obtain a reliable fixed-point location.
On the other hand the $L$ cluster does not have
such a redundancy,
and the Hamiltonian 
is given by Eq. (\ref{Hamiltonian}) with $L=4$
unambiguously.
We diagonalize these Hamiltonian matrices
numerically.
Note that we employ the conventional diagonalization method,
rather than Novotny's diagonalization method, which is utilized
in Sec. \ref{section3}; see the Appendix as well.

With use of the simulation technique developed above,
we search for the fixed point of the real-space decimation.
We survey the parameter space, regarding $\Gamma$ as a unit of energy;
namely, we set $\Gamma=1$
throughout this section. 
Thereby,
we impose the following conditions
\begin{eqnarray}
      \label{scale_invariance1}
\langle S^z_1 S^z_2 \rangle_S   &=&   \langle \tilde{S}^z_1 \tilde{S}^z_2 \rangle_L     \\
\langle S^z_1 S^z_4 \rangle_S   &=&   \langle \tilde{S}^z_1 \tilde{S}^z_4 \rangle_L     \\
\langle (S^z_1 S^z_2)^2 \rangle_S   &=&   \langle (\tilde{S}^z_1 \tilde{S}^z_2)^2 \rangle_L     \\
\langle (S^z_1 S^z_4)^2 \rangle_S   &=&   \langle (\tilde{S}^z_1 \tilde{S}^z_4)^2 \rangle_L     \\
      \label{scale_invariance5}
\langle (S^z_1)^2 \rangle_S   &=&   \langle (\tilde{S}^z_1)^2 \rangle_L    
           ,
\end{eqnarray}
as a scale-invariance criterion.
The arrangement of the spin variables 
is shown in Fig. \ref{figure1}.
the symbol $\langle \dots \rangle_{S,L}$ denotes the ground-state average
for the respective clusters.
The above equations set the scale-invariance condition
as to the correlation functions \cite{Swendsen82}.
In order to solve the non-linear equations (\ref{scale_invariance1})-(\ref{scale_invariance5}),
we utilized the Newton method,
and arrived at
the fixed point, Eq. (\ref{fixed_point});
the last digits may be uncertain because of the round off errors.
In the next section, we present the simulation result around the fixed point 
(\ref{fixed_point}).
As claimed in Ref. \cite{Wegner},
the decimation scheme based on the correlation function(s) is problematic.
In our approach, the quantitative analyses of critical exponents
are made subsequently by the finite-size scaling.

Last, we argue the validity of the fixed point (\ref{fixed_point}),
and the boundary condition $b=0.75$.
In Sec. \ref{section3_1}, we estimate $\Gamma_c=1.0007(17)$
via the finite-size-scaling analysis.
Clearly, this result is consistent with $\Gamma=1$ postulated in this section
(decimation analysis).
As a matter of fact,
the above real-space decimation contains several sources of errors,
such as the restricted system size and
a fundamental difficulty 
as to the determination of fixed point \cite{Jones78}.
Here, we make use of the redundant parameter $b$,
aiming to
control these errors
in a rather {\it ad hoc} manner.


\section{\label{section3}Numerical results}

In this section, we present the simulation result for
the $(2+1)$-dimensional Ising model (\ref{Hamiltonian}).
The coupling-constant parameters are set to the scaling-invariant point
(\ref{fixed_point}), 
at which the scaling behavior improves.
(We stress that the scale-invariant analysis of Sec. \ref{section2} is simply a preliminary one,
and it is no longer used.)
We employ Novotny's numerical diagonalization method;
the technical details are addressed in the Appendix.
Owing to this method,
we treat a variety of system sizes $N=6,8,\dots,20$
($N$ is the number of spins); 
note that conventionally the system size is restricted within $N=4,9,16,\dots$.
The linear dimension $L$ of the cluster is given by
\begin{equation}
L=\sqrt{N}            ,
\end{equation}
because the $N$ spins constitute a rectangular cluster.

Conventionally, the $d=3$ criticality has been studied
by means of the Monte Carlo method (rather than the diagonalization method).
The error of the Monte Carlo simulation comes from a purely statistical origin.
On the one hand, the diagonalization result is free from such a statistical error,
and the data analysis is not straightforward.
In the present paper,
we perform 
two independent analyses for each critical exponent,
aiming to appreciate possible systematic errors.

\subsection{\label{section3_1}
Critical point $\Gamma_c$}

In this section, we provide evidence of the
$\Gamma$-driven phase transition with the other coupling constants
set to Eq. (\ref{fixed_point}).

In Fig. \ref{figure2},
we plot the scaled energy gap $L \Delta E$
for various $\Gamma$, and $N=6,8,\dots,20$.
According to the finite-size-scaling theory,
the scaled energy gap $L \Delta E$ should be invariant at the critical point.
Indeed, we observe an onset of the $\Gamma$-driven phase transition around $\Gamma \sim 1$.

In Fig. \ref{figure3},
we plot the approximate transition point $\Gamma_c(L_1,L_2)$
for $[2/(L_1+L_2)]^3$ with $6\le N_1 < N_2 \le 20$ and $L_{1,2}=\sqrt{N_{1,2}}$;
the validity of the $1/L^3$-extrapolation scheme
(abscissa scale) is considered at the end of this section.
The approximate transition point $\Gamma_c(L_1,L_2)$ denotes
a scale-invariant point with respect to a pair of system sizes $(L_1,L_2)$.
Namely, 
according to the phenomenological renormalization
\cite{Nightingale76},
the approximate transition point satisfies the equation
\begin{equation}
\label{critical_point}
L_1 \Delta E(L_1)|_{\Gamma=\Gamma_c(L_1,L_2)} = 
L_2 \Delta E(L_2)|_{\Gamma=\Gamma_c(L_1,L_2)}
          .
\end{equation}
The least-squares fit to the data of Fig. \ref{figure3} yields 
$\Gamma_c=1.0007(17)$
in the thermodynamic limit $L \to \infty$.
As to this estimate, 
$\Gamma_c$ (\ref{critical_point}),
possible systematic errors are not appreciated,
and the amount of error margin is spurious.
Because the estimate $\Gamma_c$ is not used in the subsequent analyses
of critical exponents,
we do not go into further details; rather, $\Gamma_c(L_1,L_s)$ is used.

We mention a few remarks.
First,
we consider the abscissa scale, $1/L^3$, utilized in Fig. \ref{figure3}.
Naively, the scaling theory predicts that the dominant corrections to $\Gamma_c$ 
should scale like $1/L^{\omega+1/\nu}$ \cite{Binder81}
with $\omega=0.821(5)$ and $1/\nu=1.5868(3)$ \cite{Deng03}.
On the one hand, in our simulation,
such dominant corrections should be suppressed by adjusting the coupling constants 
to Eq. (\ref{fixed_point}).
The convergence to the thermodynamic limit may be accelerated.
(For extremely large system sizes,
the singularity $1/L^{\omega+1/\nu}$ may emerge.)
Hence, in Fig. \ref{figure3}, we set the abscissa scale to $1/L^3$.
Second, we argue a consistency between the finite-size scaling and the real-space decimation
(Sec. \ref{section2}).
In Sec. \ref{section2}, we made a fixed-point analysis, regarding 
$\Gamma$ as a unit of energy; namely, we set
$\Gamma=1$.
This proposition is quite consistent with the above simulation result,
$\Gamma_c=1.0007(17)$,
justifying the fixed-point analysis
in Sec. \ref{section2}.
In other words, around the fixed point,
corrections to scaling may cancel out satisfactorily.
Third,
the shaky character of Fig. \ref{figure3} is an artifact due to the cluster size;
as the cluster size deviates from the quadratic number 
($N=9,16,25$),
the shaky character becomes amplified. 
This feature is intrinsic to 
Novotny's diagonalization method.
(A bump of Fig. \ref{figure3}
should be attributed to this character.)

\subsection{Critical exponent $\eta$}

In Sec. \ref{section3_1}, we observed an onset of the $\Gamma$-driven phase transition.
In this section, we calculate the critical exponent $\eta$, based on the finite-size scaling.

In Fig. \ref{figure4},
we plot the approximate critical exponent
\begin{equation}
\label{eta1}
\eta(L_1,L_2)=
   -2     \ln [m(L_1)/m(L_2)]|_{\Gamma=\Gamma_c(L_1,L_2)}
               / \ln (L_1/L_2)    -1
,
\end{equation}
for $[2/(L_1+L_2)]^2$ with $6 \le N_1 < N_2 \le 20$
($L_{1,2}=\sqrt{N_{1,2}}$);
afterward, we consider the abscissa scale, $1/L^2$.
Here, the magnetization $m$ is given by 
$m=( \langle g | M^2 | g \rangle )^{1/2}/N$
with the ground state $|g \rangle$   and the total magnetization 
$M=\sum_{i=1}^N S^z_i$.
The least-squares fit to these data yields $\eta=0.03088(26)$
in the thermodynamic limit $L \to \infty$.

In Fig. \ref{figure5},
we plot the approximate critical exponent
\begin{equation}
\label{eta2}
\eta(L_1,L_2)=
-
  \ln [\chi(L_1)/\chi(L_2)]|_{\Gamma=\Gamma_c(L_1,L_2)}
               / \ln (L_1,L_2)
  +2 
,
\end{equation}
for $[2/(L_1+L_2)]^2$ with $6 \le N_1 < N_2 \le 20$.
Here, the susceptibility $\chi$ is given by the resolvent form
\begin{equation}
\chi= \frac{1}{N} \langle g | M \frac{1}{{\cal H}-E_g}M|g\rangle      ,
\end{equation}
with 
the ground state energy $E_g$.
The resolvent form is readily calculated with use of the continued-fraction method
\cite{Gagliano87}.
The least-squares fit to the data in Fig. \ref{figure5}
yields $\eta=0.03676(19)$
in the thermodynamic limit $L \to \infty$.
Considering the difference between the results of Figs. \ref{figure4} and \ref{figure5} 
as an error indicator, we estimate the critical exponent as
\begin{equation}
\eta = 0.034(4)
            .
\end{equation}
The index immediately yields the critical exponent
\begin{equation}
\label{betawnu}
\beta/\nu = 0.517(2)
,
\end{equation}
through the scaling relation.

We address a number of remarks.
First,
we consider the error margin of the critical exponent.
The result $\eta=0.03676(19)$ (Fig. \ref{figure5}) is in remarkable agreement with that of
recent Monte Carlo, $\eta=0.0368(2)$ \cite{Deng03}.
In order to appreciate the error margin properly,
one observation would not suffice, because
there should exist an error (systematic deviation)
that cannot be captured by a
statistical (least-squares fit) analysis.
Therefore, we take into account another
independent simulation result $\eta=0.03088(26)$
(Fig. \ref{figure4}).
(In general, the reliability of $N$-independent observations
is given by $\sigma/\sqrt{N-1}$ with the standard deviation
$\sigma$ of each observation.)
That is, the deviation of the results Figs. \ref{figure4} and \ref{figure5}
yields an indicator of the error margin.
Hence, we arrive at the above conclusion
with the error margin, 
which covers
these independent observations.
Second, we argue the abscissa scale $1/L^2$ utilized in Figs. \ref{figure4} and \ref{figure5}.
In Ref. \cite{Nishiyama08} (see, in particular, Fig. 3 and Sec. IIB),
it was claimed that the abscissa scale
$1/L^2$ yields a consistent result; namely, 
results via two independent analyses coincide.
Hence, we set the abscissa scale to $1/L^2$ in Figs. \ref{figure4} and \ref{figure5}.
Naively, the scaling theory predicts that the leading corrections to the critical exponent
should scale like $1/L^\omega$ with $\omega=0.821(5)$ \cite{Deng03}.
On the other hand, the convergence should be accelerated more than the naively expected one,
because such dominant corrections are truncated by adjusting the coupling-constant parameters
to Eq. (\ref{fixed_point}).
In this sense, an accelerated convergence $1/L^2$ is a reasonable one,
at least, within the range of system sizes
tractable with the diagonalization method.


\subsection{Critical exponent $\nu$}

In this section, we calculate the critical exponent $\nu$.

In Fig. \ref{figure6},
we plot the approximate critical exponent
\begin{equation}
\label{nu1}
1/\nu= 
\frac{ \ln(  \partial_\Gamma e(L_1) /\partial_\Gamma e(L_2) )  |_{\Gamma=\Gamma_c(L_1,L_2)}  }   
   { 2 \ln(L_1/L_2) }
     +\frac{3}{2}          ,
\end{equation}
for $[2/(L_1+L_2)]^2$ with $6 \le N_1<N_2\le 20$.
Here, the internal energy, $e$, is given by the off-diagonal part of the Hamiltonian
(namely, $e= - \langle g | \Gamma \sum_{j=1}^N S^x_i | g \rangle /N $),
which turns out to exhibit little size dependence.
The least-squares fit to these data yields
$1/ \nu=1.58765(25)$ in the thermodynamic limit $L \to \infty$.
In order to obtain the exponent $1/\nu$, we have to take a $\Gamma$ derivative
with respect to a certain quantity. In the preliminary stage, we surveyed
a variety of quantities such as the internal energy set tentatively to
$e=$[diagonal part of Eq. (\ref{Hamiltonian})]$/N$,
excitation gap, and susceptibility.
As a consequence, we arrived at the conclusion
that the above-mentioned clue (\ref{nu1}) exhibits
little size dependence.
Similar tendency is observed \cite{Nishiyama99} for the case of $S=1/2$,
namely, the conventional transverse-field Ising model.
In the case of $S=1/2$, via the same scheme, we obtained
a tentative result $1/\nu \approx 1.6$ even without making
any extrapolation.
This fact indicates that the $\Gamma$ derivative
$\partial_\Gamma e=C_0 + A_0 L^{2/\nu-3}$ has a negligible constant term
$C_0$, and the singular term $A_0$ is dominating.
In the analysis of Fig. \ref{figure6}, we made use of this character.


In Fig. \ref{figure7},
we plot the approximate critical exponent
\begin{equation}
\label{nu2}
1/\nu-\beta/\nu= \ln(  \partial_\Gamma m(L_1) /\partial_\Gamma m(L_2) |_{\Gamma=\Gamma_c(L_1,L_2)}
                  )/\ln(L_1/L_2) 
      ,
\end{equation}
for $[2/(L_1+L_2)]^2$ with $6 \le N_1<N_2\le 20$.
The least-squares fit to these data yields
$1/\nu-\beta/\nu=1.0676(55)$ in the thermodynamic limit $L \to \infty$.
Combining this result with Eq. (\ref{betawnu}),
we arrive at $1/\nu=1.5846(59)$.
Because it is calculated indirectly, it might have uncontrolled deviation.
Nevertheless, the consistency 
to the above result (Fig. \ref{figure6}) seems to be satisfactory.
Counting these independent results, we estimate
\begin{equation}
1/\nu = 1.586 (6)
         ,
\end{equation}
with the error margin coming from the latter one.

\subsection{Simulation at $(J,J',J_4,J_4',D)=(0.6,0,0,0,0)$}

In the above,
we simulated the transverse-field Ising model, Eq. (\ref{Hamiltonian}),
with the finely tuned coupling constants,
Eq. (\ref{fixed_point}).
As a comparison, in this section, we provide
a result for a model without the extended interactions;
namely, we set
$(J,J,J,J,D)=(0.6,0,0,0,0)$ tentatively.

In Fig. \ref{figure8},
we plot the scaled energy gap $L \Delta E$ for various $\Gamma$
and $N=6,8,\dots,20$. 
We observe an onset of the $\Gamma$-driven phase transition around $\Gamma \sim 1$.
However, the location of critical point appears to be unclear, as compared to that of Fig. \ref{figure2}.
This result demonstrates clearly that the finely-tuned coupling constants
lead to an elimination of finite-size corrections.

\section{\label{section4}Summary and discussions}

\begin{figure}
\includegraphics[width=100mm]{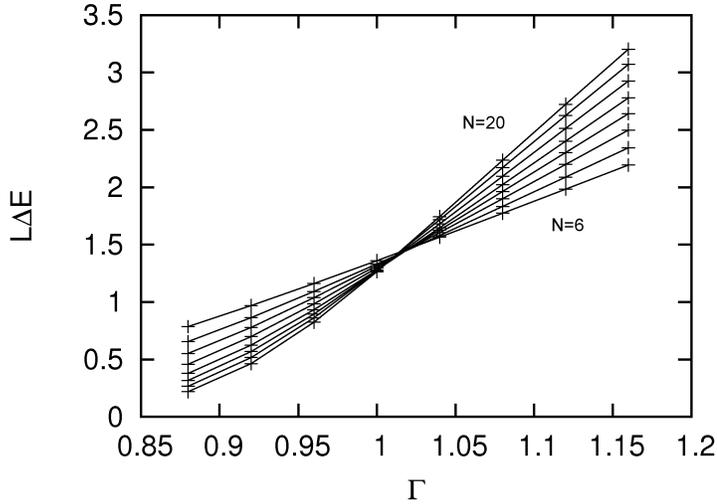}%
\caption{\label{figure2}
Scaled energy gap $L \Delta E$ is plotted for various $D$ and 
$N=6,8,\dots,20$ ($L=\sqrt{N}$);
note that we survey the $\Gamma$-driven phase transition with the other interactions 
$(J,J',J_4,J_4',D)$ adjusted
to a fixed point (\ref{fixed_point}).
We observe a clear indication of the $\Gamma$-driven transition around $\Gamma\sim 1$.
Clearly, the scaling behavior is improved as compared to that of the conventional 
transverse-field Ising model without the extended interactions (Fig. \ref{figure8}).
}
\end{figure}

\begin{figure}
\includegraphics[width=100mm]{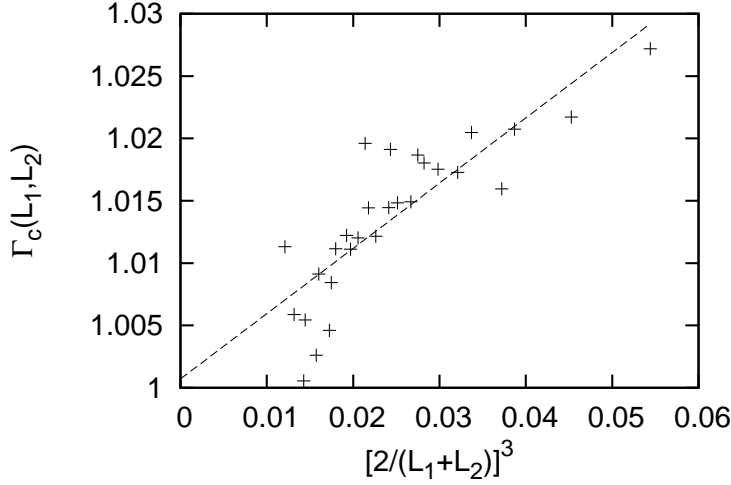}%
\caption{\label{figure3}
The approximate critical point $\Gamma_c$ (\ref{critical_point}) is plotted
for $[2/(L_1+L_2)]^3$ with $6 \le N_1<N_2 \le 20$
($L_{1,2}=\sqrt{N_{1,2}}$).
The least-squares fit to these data yields $\Gamma_c=1.0007(17)$ in the thermodynamic limit $L\to \infty$.
}
\end{figure}

\begin{figure}
\includegraphics[width=100mm]{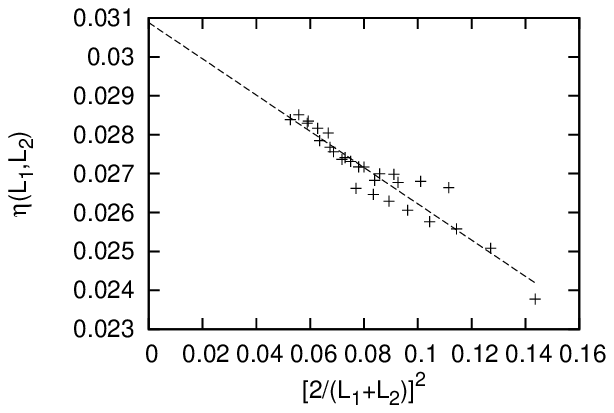}%
\caption{\label{figure4}
The approximate critical exponent $\eta(L_1,L_2)$ (\ref{eta1}) is plotted for 
$[2/(L_1+L_2)]^2$ with 
$6 \le N_1<N_2\le 20$ 
($L_{1,2}=\sqrt{N_{1,2}}$).
The least-squares fit to these data yields $\eta=0.03088(26)$ in the thermodynamic limit $L\to \infty$.
}
\end{figure}

\begin{figure}
\includegraphics[width=100mm]{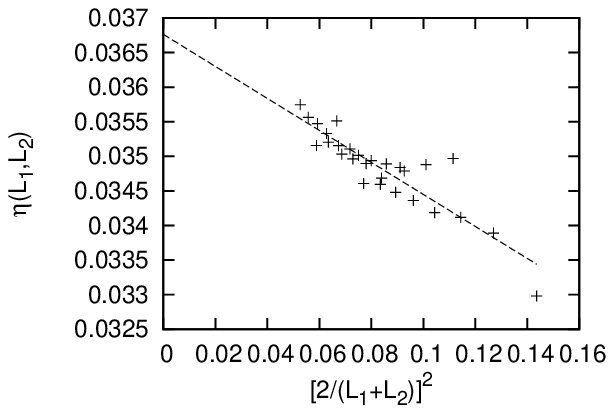}%
\caption{\label{figure5}
The approximate critical exponent $\eta(L_1,L_2)$ (\ref{eta2}) is plotted for 
$[2/(L_1+L_2)]^2$ with 
$6 \le N_1<N_2\le 20$ 
($L_{1,2}=\sqrt{N_{1,2}}$).
The least-squares fit to these data yields $\eta=0.03676(19)$ in the thermodynamic limit $L\to \infty$.
}
\end{figure}

\begin{figure}
\includegraphics[width=100mm]{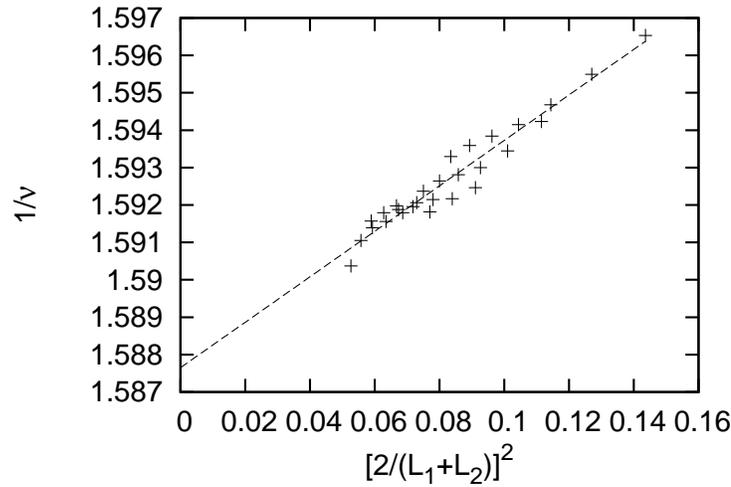}%
\caption{\label{figure6}
The approximate critical exponent $1/\nu$ (\ref{nu1}) is plotted for 
$[2/(L_1+L_2)]^2$ with 
$6 \le N_1<N_2\le 20$ 
($L_{1,2}=\sqrt{N_{1,2}}$).
The least-squares fit to these data yields $1/\nu=1.58765(25)$ in the thermodynamic limit $L\to \infty$.
}
\end{figure}

\begin{figure}
\includegraphics[width=100mm]{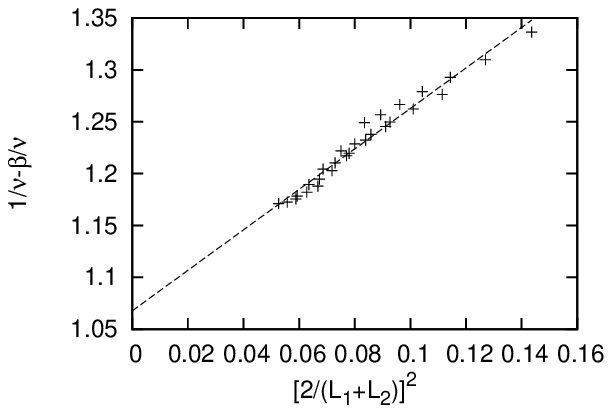}%
\caption{\label{figure7}
The approximate critical exponent $1/\nu-\beta/\nu$ (\ref{nu2}) is plotted for 
$[2/(L_1+L_2)]^2$ with 
$6 \le N_1<N_2\le 20$ 
($L_{1,2}=\sqrt{N_{1,2}}$).
The least-squares fit to these data yields $1/\nu-\beta/\nu=1.0676(55)$ in the thermodynamic limit $L\to \infty$.
}
\end{figure}

\begin{figure}
\includegraphics[width=100mm]{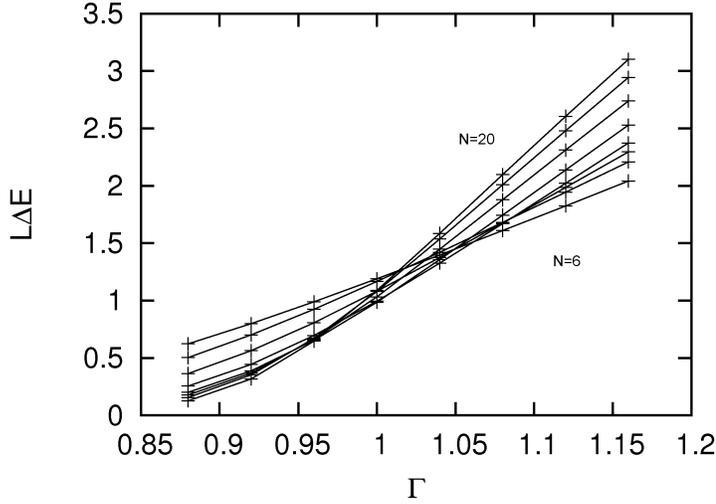}%
\caption{\label{figure8}
Tentatively, we turned off the extended interactions,
$(J,J',J_4,J_4',D)=(0.6,0,0,0,0)$,
and calculated the scaled energy gap
$L\Delta E$ for various $\Gamma$ and $N=6,8,\dots,20$ ($L=\sqrt{N}$).
We notice that the data are scattered as compared to those of Fig. \ref{figure2}.
}
\end{figure}

The $(2+1)$-dimensional $S=1$ transverse-field Ising model (\ref{Hamiltonian})
was investigated numerically.
Unlike the conventional $S=1/2$ spin model,
the $S=1$ model allows us to incorporate a variety of interactions,
$(J,J',J_4,J_4',D)$.
By adjusting these interactions to a scale-invariant point (\ref{fixed_point}),
we attain substantial elimination of scaling errors;
see Figs. \ref{figure2} and \ref{figure8}.
As a result, we estimate the critical indices as 
$\eta=0.034(4)$ and 
$1/\nu=1.586(6)$.
Through the scaling relations,
we arrive at
\begin{equation}
\label{abc}
(\alpha,\beta,\gamma)=[
0.1084(72),
0.3260(18),
1.2396(53)] .
\end{equation}

We overview related studies.
As mentioned in the Introduction, 
a set of indices $[0.1265(84),0.3304(48),1.213(11)]$ was obtained
in our preliminary survey \cite{Nishiyama06}
as for the Ising model with the next-nearest-neighbor and plaquette-four-spin interactions.
The extension of interactions, Eq. (\ref{Hamiltonian}), leads to an improvement as to the indices
(\ref{abc}).
Moreover, in the present study,
we manage two independent analyses of  each index.
Hence, it is expected that the error margins of Eq. (\ref{abc})
would be appreciated properly.
A numerical diagonalization result 
$[0.1144(24),0.3281(13),1.2319(65)]$ \cite{Hamer00}
was evaluated
for a large $6\times 6$ cluster (without the extended interactions) for $S=1/2$;
a diagonalization of a $7 \times 7$ cluster ($N=49$)
far exceeds a limitation
of available computer resources.
The present scheme (see the Appendix) 
admits us to enlarge $N$ linearly (for example, $N=6,8,10,\dots$),
and an extension of $N$ might not be so computationally demanding.
The series-expansion method yields the following indices
via
the
high-temperature expansion
$[0.1096(5),0.32653(10),1.2373(2)]$ \cite{Campostrini02}
as well as the
field-theoretical analyses,
$[0.109(4),0.3258(14),1.2396(13)]$
\cite{Guida98}
and
$[0.1091(24),0.3257(5),1.2403(8)]$
\cite{Jasch01}.
Recent Monte Carlo results are
$[0.1109$ $(15),$ $0.3262(4),1.2366(15)]$ \cite{Hasenbusch01}
and
$[0.10940(36),0.326695(88),1.23721(27)]$ \cite{Deng03},
The latter result, which has considerably small error margins,
was estimated by surveying a variety of lattice structures.
Actually, the character of scaling corrections depends 
on the respective lattice structures.
Hence, it is sensible to count various lattices
in order to reduce and certify the extrapolation errors;
in the present paper, we carry out
two independent analyses for each critical exponent.
At present,
such 
{\it a posteriori} consideration cannot be omitted.
In this sense, further consideration of scaling corrections
would be desirable in order to estimate the critical indices in a 
fully controlled manner.

\appendix

\section{Numerical method: Novotny's diagonalization scheme}

In Sec. \ref{section3},
we simulated the two-dimensional transverse-field Ising model (\ref{Hamiltonian})
with Novotny's diagonalization method \cite{Novotny90,Novotny92}.
In this Appendix, we explain the simulation algorithm.

To begin with, we outline the simulation algorithm.
The spins constitute a one-dimensional ($d=1$) alignment rather than a $d=2$ cluster.
The dimensionality is lifted to $d=2$ effectively by introducing the long-range interactions
with the (average) distance $\sqrt{N}$.
Because of this geometrical character, one is able to construct a cluster with an arbitrary number
of spins $N=6,8,\dots,20$;
note that conventionally, the system sizes are restricted within $N=4,9,16,\dots$.

Originally, Novotny's method was developed for the classical Ising model.
Meanwhile, it was extended (formulated) to adopt the transverse magnetic field
\cite{Nishiyama07a};
see the Appendix of Ref. \cite{Nishiyama07b} as well.
In the present paper, we resort to the formalism
addressed in Ref. \cite{Nishiyama07b}.
In the following, we summarize the (almost trivial) modifications
in order to extend the spin magnitude to $S=1$ from $S=1/2$.
First, the notation for the spin operator, $\sigma^{x,y,z}_i$ (Pauli matrices),
has to be replaced with $S^{x,y,z}_i$.
Correspondingly,
the
Hilbert-space basis [Eq. (A1) of Ref. \cite{Nishiyama07b}]
is represented by the expression
$ | \{ S_i \}\rangle$ with $i=1,2,\dots,N$ and $S_i=-1,0,1$.
Second,
the plaquette interaction [summand of Eq. (A5) in Ref. \cite{Nishiyama07b}] has to be replaced with
an extended one
\begin{eqnarray}
\langle \{ S_i \} | T | \{ T_i \} \rangle      &=&
    \sum_{j=1}^N
(
-\frac{J}{2}(S_jS_{j+1} +T_jT_{j+1}+S_jT_j+S_{j+1}T_{j+1}) \nonumber \\
& & -J'(S_j T_{j+1}+S_{j+1}T_j)                    \nonumber \\
  & & -\frac{J_4}{2}((S_jS_{j+1})^2 +(T_jT_{j+1})^2+(S_jT_j)^2+(S_{j+1}T_{j+1})^2)  \nonumber \\
& & -J_4'((S_j T_{j+1})^2+(S_{j+1}T_j)^2)       \nonumber   \\
& &   +\frac{D}{2} (S_j^2 +T_j^2)
)
.
\end{eqnarray}

Technically, the above modifications suffice
for the simulation of the Hamiltonian (\ref{Hamiltonian}).
In the following, we provide a supplementary remark
concerning the translational operator $P$; see Eq. (A4) of Ref. \cite{Nishiyama07b}.
We calculate the matrix element of the operator $P^v$
with $v=\sqrt{N}$,
based on the expansion
\begin{equation}
\label{expansion_P}
\langle  \{ S_i\} | P^v | \{ T_i\}\rangle
              = \sum_k \sum_{\Psi_k}
        \langle \{ S_i \} | \Psi_k \rangle
               \exp(  \frac{2\pi kv}{N}i )
              \langle \Psi_k | \{ T_i \} \rangle
  ,
\end{equation}
with the intermediate states 
$|\Psi_k \rangle$ and an integer index $k$ specifying a Brillouin zone
(reciprocal space);
note that
the reciprocal space $\{ k \}$ is suitable for representing $P$.
The point is that
each Brillouin zone is no longer equivalent 
because of the incommensurate oscillating factor 
$\exp (2\pi k vi/N)$ in Eq. (\ref{expansion_P}).
The range of Brillouin zone of the present simulation is $-N/2 \le k \le N/2$.
More specifically, the intermediate sum, $\sum_k$, in 
Eq. (\ref{expansion_P})
is expanded into
\begin{equation}
\sum_k a_k=\frac{1}{2} a_{-N/2}+a_{-N/2+1}+a_{-N/2+2}+\dots+a_{N/2-1}+\frac{1}{2}a_{N/2}
 ,
\end{equation}
for a summand $a_k$.
The factor $1/2$ compensates the duplicated sum at the zone boundaries.




\begin{thebibliography}{99}

\bibitem{Chen82}J.H. Chen, M.E. Fisher, and B.G. Nickel,
Phys. Rev. Lett. {\bf 48} (1982) 630.
\bibitem{Symanzik83a}K. Symanzik, Nucl. Phys. B {\bf 226} (1983) 187.
\bibitem{Symanzik83b}K. Symanzik, Nucl. Phys. B {\bf 226} (1983) 205.


\bibitem{Hasenfratz94}P. Hasenfratz and F. Niedermayer,
Nucl. Phys. B {\bf 414} (1994) 785 (1994).
\bibitem{Hasenfratz98}P. Hasenfratz, Prog. Theor. Phys. Suppl. {\bf 131} (1998) 189.

\bibitem{Blote95}H.W.J. Bl\"ote, E. Luijten, and J.R. Heringa,
J. Phys. A: Math. Gen. {\bf 28} (1995) 6289.


\bibitem{Fernandez94}L.A. Fern\'andez, A. Mu\~noz Sudupe, J.J. Ruiz-Lorenzo,
and A. Taranc\'on, Phys. Rev. D {\bf 50} (1994) 5935.





\bibitem{Ballesteros98}
H.G. Ballesteros, L.A. Fern\'andez, V. Mart\'in-Mayor, and
A. Mu\~noz Sudupe,
Phys. Lett. B {\bf 441} (1998) 330.
\bibitem{Hasenbusch99}
M. Hasenbusch, K. Pinn, and S. Vinti,
Phys. Rev. B {\bf 59} (1999) 11471.
\bibitem{Hasenbusch00}
M. Hasenbusch and T. Torok,
Nucl. Phys. B (Proc. Suppl.) {\bf 83-4} (2000) 694.


\bibitem{Nishiyama06}
Y. Nishiyama,
Phys. Rev. E {\bf 74} (2006) 016120.


\bibitem{Deng03}
Y. Deng and H.W.J. Bl\"ote,
Phys. Rev. E {\bf 68} (2003) 036125.



%
\bibitem{Swendsen82}
R.H. Swendsen, in
{\it Real-Space Renormalization},
edited by T.W. Burkhardt and J.M.J. van Leeuwen
(Springer-Verlag, Berlin, 1982). 

\bibitem{Wegner}
F.J. Wegner,
{\it Phase Transitions and Critical Phenomena 6},
edited by
C. Domb and M.S. Green
(Academic Press, London and New York, 1976), p. 29.

\bibitem{Jones78}
G.L. Jones, J. Stat. Phys. {\bf 19}, 417 (1978).
\bibitem{Nightingale76}
M.P. Nightingale, Physica {\bf 83A}, 561 (1976).


\bibitem{Binder81}
K. Binder, Z. Phys. B: Condens. Matter {\bf 43} (1981) 119.

\bibitem{Nishiyama08}
Y. Nishiyama, Phys. Rev. E {\bf 77} (2008) 051112.




\bibitem{Gagliano87}
E. R. Gagliano and C. A. Balseiro,
Phys. Rev. Lett. {\bf 59} (1987) 2999.


\bibitem{Hamer00}
C.J. Hamer, J. Phys. A {\bf 33} (2000) 6683.


\bibitem{Nishiyama99}
Y. Nishiyama, in preparation.


\bibitem{Campostrini02}
M. Campostrini, A. Pelissetto, P. Rossi, and E. Vicari, Phys. Rev. E {\bf 65} (2002) 066127.



\bibitem{Guida98}
R. Guida and J. Zinn-Justin, J. Phys. A {\bf 31} (1998) 8103.

\bibitem{Jasch01}
F. Jasch and H. Kleinert, J. Math. Phys. {\bf 42} (2001) 52.


\bibitem{Hasenbusch01}
M. Hasenbusch, Int J. Mod. Phys. C {\bf 12} (2001) 911.



%
\bibitem{Novotny90}M.A. Novotny, J. Appl. Phys. {\bf 67} (1990) 5448.
\bibitem{Novotny92}M.A. Novotny, Phys. Rev. B {\bf 46} (1992) 2939.







\bibitem{Nishiyama07a}
Y. Nishiyama, Phys. Rev. E {\bf 75} (2007) 011106.

\bibitem{Nishiyama07b}
Y. Nishiyama, Phys. Rev. E {\bf 75} (2007) 051116.


\end{thebibliography}







\end{document}